\documentclass[10pt]{article}
\usepackage{graphicx}
\usepackage{amsmath}
\usepackage{amssymb}
\usepackage{caption2}
\begin{document}
\begin{center}
{\large {\bf \sc{   Analysis of the  $X_0(2900)$ as the scalar tetraquark state   via the  QCD sum rules
  }}} \\[2mm]
Zhi-Gang  Wang \footnote{E-mail: zgwang@aliyun.com.  }   \\
 Department of Physics, North China Electric Power University, Baoding 071003, P. R. China
\end{center}

\begin{abstract}
In this article, we study the axialvector-diquark-axialvector-antidiquark ($AA$) type and scalar-diquark-scalar-antidiquark ($SS$) type fully open flavor $cs\bar{u}\bar{d}$ tetraquark states with the spin-parity $J^P={0}^+$ via the QCD sum rules. The predicted masses $M_{AA}=2.91\pm0.12\,\rm{GeV}$ and $M_{SS}=3.05\pm0.10\,\rm{GeV}$  support assigning the $X_0(2900)$ to be the $AA$-type scalar  $cs\bar{u}\bar{d}$ tetraquark state.
\end{abstract}



\section{Introduction}
Recently, the LHCb collaboration reported a narrow peak in the $D^- K^+$
invariant mass spectrum in the decays $B^\pm\to D^+ D^- K^\pm$ with the statistical significance much greater  than $ 5\sigma$ \cite{LHCbTalk:2020-1,LHCbTalk:2020-2}.  The peak has
been parameterized in terms of two Breit-Wigner resonances:
\begin{eqnarray}
X_0(2900) &:& J^P=0^+,~M_0=2866\pm7~{\rm MeV},~~\Gamma_0=\phantom{1}57\pm13~{\rm MeV}~;\\
X_1(2900) &:& J^P=1^-,~M_1=2904\pm5~{\rm MeV},~~\Gamma_1=110\pm12~{\rm MeV}~.
\end{eqnarray}
This is the first exotic hadron with fully open flavor,  the valence quarks or the constituent quarks are $cs\bar{u}\bar{d}$ \cite{LHCbTalk:2020-1,LHCbTalk:2020-2}. In Ref.\cite{Karliner-2900}, Karliner and Rosner assign the narrow peak to be the scalar-diquark-scalar-antidiquark  type tetraquark state with the spin-parity $J^P=0^+$.
Subsequently, other assignments are proposed, such as the $D^*\bar{K}^*$ molecular state \cite{DK-molecule-1,DK-molecule-2,DK-molecule-3}, the radial excited tetraquark state or orbitally excited tetraquark state \cite{Radial-excite}, the triangle singularity \cite{Triangle}, the scalar tetraquark state \cite{JRZhang-2900}, non scalar tetraquark state \cite{QFLu}.

In 2019, the BESIII  collaboration  explored  the process $J/\psi \to \phi \eta \eta^\prime$ and observed a structure $X$ in  the $\phi\eta^\prime$ mass spectrum \cite{BES-2000}. The fitted mass and width are $M_X=(2002.1\pm 27.5 \pm 15.0)\,\rm{MeV}$ and $\Gamma_X=(129 \pm 17 \pm 7)\,\rm{MeV}$ respectively with the assignment  $J^P=1^-$,  while the fitted mass and width are $M_X=(2062.8 \pm 13.1 \pm 4.2)
\,\rm{MeV}$ and $\Gamma_X=(177 \pm 36 \pm 20)\,\rm{MeV}$ respectively with  the assignment  $J^P=1^+$.
In Ref.\cite{WangAdvHigh}, we study the axialvector-diquark-axialvector-antidiquark type  scalar, axialvector, tensor and vector  $ss\bar{s}\bar{s}/qq\bar{q}\bar{q}$  tetraquark states with the QCD sum rules in a systematic way.  The predicted mass  $M_{X}=2.08\pm0.12\,\rm{GeV}$ for the axialvector tetraquark state  supports assigning the new structure $X(2060)$ from the BESIII  collaboration  to be a $ss\bar{s}\bar{s}$ tetraquark state with the spin-parity-charge-conjugation $J^{PC}=1^{+-}$.  In Ref.\cite{Wang-Hidden-charm}, we  construct various  scalar, axialvector and tensor tetraquark currents to study the  mass spectrum of the ground state hidden-charm tetraquark states  with
the QCD sum rules in a comprehensive way, and revisit the assignments of the $X$, $Y$, $Z$ states, such as  $X(3860)$, $X(3872)$, $X(3915)$,  $X(3940)$, $X(4160)$, $Z_c(3900)$, $Z_c(4020)$, $Z_c(4050)$, $Z_c(4055)$, $Z_c(4100)$, $Z_c(4200)$, $Z_c(4250)$, $Z_c(4430)$, $Z_c(4600)$, etc  in a consistent way. For the axialvector-diquark-axialvector-antidiquark type ($AA$-type)  scalar tetraquark states, we obtain the masses \cite{WangAdvHigh,Wang-Hidden-charm},
\begin{eqnarray}
M_{qq\bar{q}\bar{q}}&=&1.86\pm0.11\, \rm{GeV}\, ,\nonumber\\
M_{ss\bar{s}\bar{s}}&=&2.08\pm0.13\, \rm{GeV}\, ,\nonumber\\
M_{cq\bar{c}\bar{q}}&=&3.95\pm 0.09\, \rm{GeV}\, .
\end{eqnarray}
Now we can estimate the mass of the $AA$-type $cs\bar{u}\bar{d}$ tetraquark state crudely,
\begin{eqnarray}
M_{cs\bar{u}\bar{d}}&=&\frac{M_{qq\bar{q}\bar{q}}+M_{ss\bar{s}\bar{s}}+2M_{cq\bar{c}\bar{q}}}{4}=2.96\pm0.11\,\rm{GeV}\, ,
\end{eqnarray}
which is consistent with the mass of the $X_0(2900)$ within uncertainties.

In this article, we construct the scalar-diquark-scalar-antidiquark type ($SS$-type) and axialvector-diquark-axialvector-antidiquark type ($AA$-type) scalar  currents to study the masses of the $cs\bar{u}\bar{d}$ tetraquark states with the QCD sum rules in details and explore the possible assignment of the $X_0(2900)$ as the scalar tetraquark state.

The article is arranged as follows:  we obtain  the QCD sum rules for the masses and pole residues of  the
 scalar tetraquark  states in Sect.2;  in Sect.3, we present the numerical results and discussions; and Sect.4 is reserved for our
conclusion.

\section{QCD sum rules for  the scalar tetraquark  states}

Firstly, we write down  the two-point correlation functions  $\Pi(p^2)$  in the QCD sum rules,
\begin{eqnarray}
\Pi(p^2)&=&i\int d^4x e^{ip \cdot x} \langle0|T\left\{J(x)\bar{J}(0)\right\}|0\rangle \, ,
\end{eqnarray}
where $J(x)=J_{AA}(x)$,  $J_{SS}(x)$,
 \begin{eqnarray}
 J_{AA}(x)&=&\varepsilon^{ijk} \varepsilon^{imn} s^T_j(x) C\gamma_{\alpha} c_k(x)\,\bar{u}_m(x) \gamma^{\alpha} C \bar{d}^T_n(x)\,  , \nonumber\\
  J_{SS}(x)&=&\varepsilon^{ijk} \varepsilon^{imn} s^T_j(x) C\gamma_5 c_k(x)\,\bar{u}_m(x) \gamma_5 C \bar{d}^T_n(x)\,  ,
\end{eqnarray}
 the $i$, $j$, $k$,  $m$ and $n$ are color indexes, the $C$ is the charge conjugation matrix.
The attractive interactions of one-gluon exchange  favor  formation of
the diquarks in  color antitriplet  \cite{One-gluon-1,One-gluon-2}. The  QCD sum rules calculations indicate that  the favored quark-quark configurations are the scalar  and axialvector  diquark states \cite{Dosch-Diquark-1989-1,Dosch-Diquark-1989-2,WangDiquark-1,WangDiquark-2,WangLDiquark}.

At the hadron  side,   we  insert  a complete set  of scalar  tetraquark states with the same quantum numbers as the current operators $J(x)$
  into the correlation functions $\Pi(p^2)$   to obtain the hadronic representation
\cite{SVZ79-1,SVZ79-2,Reinders85}. After isolating the pole terms of the lowest $cs\bar{u}\bar{d}$   tetraquark states $X_0$, we obtain the result:
\begin{eqnarray}
  \Pi(p^2) & = & \frac{\lambda_X^2}{M_X^2-p^2}    + \cdots \, ,
    \end{eqnarray}
where the pole residues $\lambda_X$ are defined by $\langle 0|J(0)|X(p)\rangle=\lambda_X$.

 Now, we briefly outline  the operator product expansion for the correlation functions $\Pi(p^2)$  in perturbative QCD. Firstly,  we contract the $u$, $d$, $s$ and $c$ quark fields in the correlation functions $\Pi(p^2)$    with Wick theorem, and obtain the result:
\begin{eqnarray}
\Pi_{AA}(p^2)&=&i\,\varepsilon^{ijk}\varepsilon^{imn}\varepsilon^{i^{\prime}j^{\prime}k^{\prime}}\varepsilon^{i^{\prime}m^{\prime}n^{\prime}}   \int d^4x\, e^{ip\cdot x} \nonumber\\
&&{\rm Tr}\left[\gamma_\mu C_{kk^\prime}(x)   \gamma_\nu C S_{jj^\prime}^T(x)C\right] {\rm Tr}\left[\gamma^\nu U_{m^\prime m}(-x)   \gamma^\mu C D_{n^\prime n }^T(-x)C\right]\, , \\
\Pi_{SS}(p^2)&=&i\,\varepsilon^{ijk}\varepsilon^{imn}\varepsilon^{i^{\prime}j^{\prime}k^{\prime}}\varepsilon^{i^{\prime}m^{\prime}n^{\prime}}   \int d^4x\, e^{ip\cdot x} \nonumber\\
&&{\rm Tr}\left[\gamma_5 C_{kk^\prime}(x)   \gamma_5 C S_{jj^\prime}^T(x)C\right] {\rm Tr}\left[\gamma_5 U_{m^\prime m}(-x)   \gamma_5 C D_{n^\prime n }^T(-x)C\right]\, ,
\end{eqnarray}
where
the $U_{ij}(x)$, $D_{ij}(x)$, $S_{ij}$ and $C_{ij}(x)$ are the full $u$, $d$, $s$ and $c$ quark propagators, respectively,
 \begin{eqnarray}
U/D_{ij}(x)&=& \frac{i\delta_{ij}\!\not\!{x}}{ 2\pi^2x^4}-\frac{\delta_{ij}\langle
\bar{q}q\rangle}{12} -\frac{\delta_{ij}x^2\langle \bar{q}g_s\sigma Gq\rangle}{192} -\frac{ig_sG^{a}_{\alpha\beta}t^a_{ij}(\!\not\!{x}
\sigma^{\alpha\beta}+\sigma^{\alpha\beta} \!\not\!{x})}{32\pi^2x^2}  \nonumber\\
&&  -\frac{1}{8}\langle\bar{q}_j\sigma^{\mu\nu}q_i \rangle \sigma_{\mu\nu}+\cdots \, ,
\end{eqnarray}
\begin{eqnarray}
S_{ij}(x)&=& \frac{i\delta_{ij}\!\not\!{x}}{ 2\pi^2x^4}
-\frac{\delta_{ij}m_s}{4\pi^2x^2}-\frac{\delta_{ij}\langle
\bar{s}s\rangle}{12} +\frac{i\delta_{ij}\!\not\!{x}m_s
\langle\bar{s}s\rangle}{48}-\frac{\delta_{ij}x^2\langle \bar{s}g_s\sigma Gs\rangle}{192}+\frac{i\delta_{ij}x^2\!\not\!{x} m_s\langle \bar{s}g_s\sigma
 Gs\rangle }{1152}\nonumber\\
&& -\frac{ig_s G^{a}_{\alpha\beta}t^a_{ij}(\!\not\!{x}
\sigma^{\alpha\beta}+\sigma^{\alpha\beta} \!\not\!{x})}{32\pi^2x^2} -\frac{i\delta_{ij}x^2\!\not\!{x}g_s^2\langle \bar{s} s\rangle^2}{7776} -\frac{\delta_{ij}x^4\langle \bar{s}s \rangle\langle g_s^2 GG\rangle}{27648}-\frac{1}{8}\langle\bar{s}_j\sigma^{\mu\nu}s_i \rangle \sigma_{\mu\nu} \nonumber\\
&&   +\cdots \, ,
\end{eqnarray}
\begin{eqnarray}
C_{ij}(x)&=&\frac{i}{(2\pi)^4}\int d^4k e^{-ik \cdot x} \left\{
\frac{\delta_{ij}}{\!\not\!{k}-m_c}
-\frac{g_sG^n_{\alpha\beta}t^n_{ij}}{4}\frac{\sigma^{\alpha\beta}(\!\not\!{k}+m_c)+(\!\not\!{k}+m_c)
\sigma^{\alpha\beta}}{(k^2-m_c^2)^2}\right.\nonumber\\
&&\left. -\frac{g_s^2 (t^at^b)_{ij} G^a_{\alpha\beta}G^b_{\mu\nu}(f^{\alpha\beta\mu\nu}+f^{\alpha\mu\beta\nu}+f^{\alpha\mu\nu\beta}) }{4(k^2-m_c^2)^5}+\cdots\right\} \, ,\nonumber\\
f^{\alpha\beta\mu\nu}&=&(\!\not\!{k}+m_c)\gamma^\alpha(\!\not\!{k}+m_c)\gamma^\beta(\!\not\!{k}+m_c)\gamma^\mu(\!\not\!{k}+m_c)\gamma^\nu(\!\not\!{k}+m_c)\, ,
\end{eqnarray}
and  $t^n=\frac{\lambda^n}{2}$, the $\lambda^n$ is the Gell-Mann matrix
\cite{Reinders85,Pascual-1984,WangHuang3900}.
We retain the terms $\langle\bar{q}_j\sigma_{\mu\nu}q_i \rangle$ and $\langle\bar{s}_j\sigma_{\mu\nu}s_i \rangle$ come from  Fierz re-ordering of the
$\langle q_i \bar{q}_j\rangle$ and $\langle s_i \bar{s}_j\rangle$  to  absorb the gluons  emitted from other quark lines to  extract the mixed condensate   $\langle\bar{q}g_s\sigma G q\rangle$ and $\langle\bar{s}g_s\sigma G s\rangle$, respectively \cite{WangHuang3900}.
Then we compute  the integrals both in the coordinate space and momentum space to obtain the correlation functions $\Pi(p^2)$. Finally, we  obtain the QCD spectral densities $\rho(s)$ at the quark level through dispersion  relation,
 \begin{eqnarray}
\rho(s)&=&{\rm lim}_{\epsilon \to 0} \frac{{\rm Im}\Pi(s+i\epsilon)}{\pi}\, .
\end{eqnarray}
  In this article, we carry out the
operator product expansion up to the vacuum condensates  of dimension-11, and
assume vacuum saturation for the  higher dimensional vacuum condensates. There are three light quark propagators and one heavy quark propagator in the correlation functions $\Pi(p^2)$, if the heavy quark line
emits a gluon and each light quark line contributes a quark-antiquark pair, we obtain a quark-gluon  operator $g_sG_{\mu\nu} \bar{q}q\bar{q}q\bar{s}s$, which is of dimension 11, and can lead to the vacuum condensates
 $\langle\bar{q}q\rangle^2\langle\bar{s}g_s\sigma Gs\rangle$ and $\langle\bar{q}q\rangle \langle\bar{s}s\rangle\langle\bar{q}g_s\sigma Gq\rangle$, we should take into account the vacuum condensate up to dimension $11$ in a consistent way. As the vacuum condensates are the vacuum expectations  of the quark-gluon operators, we take into account the quark-gluon operators of the orders $\mathcal{O}(\alpha_s^k)$ with $k\leq 1$ \cite{WangHuang3900,Wang-afsk-1,Wang-afsk-2}.

Now  we can take the quark-hadron duality below the continuum thresholds  $s_0$ and perform  the Borel transform to obtain  the   QCD sum rules:
\begin{eqnarray}\label{QCDSR}
 \lambda^{2}_{X}\exp\left( -\frac{M_{X}^2}{T^2}\right)&=& \int_{m_c^2}^{s_0}ds\, \rho(s)\exp\left( -\frac{s}{T^2}\right)\, ,
\end{eqnarray}
where the $T^2$ is the Borel parameter, $\rho(s)=\rho_{AA}(s)$, $\rho_{SS}(s)$, we neglect the explicit expressions for simplicity.

We  differentiate  Eq.\eqref{QCDSR} with respect to  $\tau=\frac{1}{T^2}$, then eliminate the
 pole residues $\lambda_{X}$ and obtain the QCD sum rules for  the masses of the scalar $cs\bar{u}\bar{d}$ tetraquark states,
 \begin{eqnarray}\label{QCDSR-mass}
 M^2_{X} &=& \frac{-\frac{d}{d\tau}\int_{m_c^2}^{s_0}ds \,\rho(s)\exp\left( -s\tau\right)}{\int_{m_c^2}^{s_0}ds\, \rho(s)\exp\left( -s\tau\right)}\,  .
\end{eqnarray}

\section{Numerical results and discussions}
We take  the standard values of the  vacuum condensates
$\langle\bar{q}q \rangle=-(0.24\pm 0.01\, \rm{GeV})^3$,  $\langle\bar{s}s \rangle=(0.8\pm0.1)\langle\bar{q}q \rangle$,
 $\langle\bar{q}g_s\sigma G q \rangle=m_0^2\langle \bar{q}q \rangle$, $\langle\bar{s}g_s\sigma G s \rangle=m_0^2\langle \bar{s}s \rangle$,
$m_0^2=(0.8 \pm 0.1)\,\rm{GeV}^2$, $\langle \frac{\alpha_s
GG}{\pi}\rangle=(0.33\,\rm{GeV})^4$    at the energy scale  $\mu=1\, \rm{GeV}$
\cite{SVZ79-1,SVZ79-2,Reinders85,ColangeloReview}, and  take the $\overline{MS}$ masses $m_{c}(m_c)=(1.275\pm0.025)\,\rm{GeV}$
 and $m_s(\mu=2\,\rm{GeV})=(0.095\pm0.005)\,\rm{GeV}$
 from the Particle Data Group \cite{PDG}.
Furthermore,  we take into account
the energy-scale dependence of  the quark condensates, mixed quark condensates and $\overline{MS}$ masses according to  the renormalization group equation \cite{Narison-mix},
 \begin{eqnarray}
 \langle\bar{q}q \rangle(\mu)&=&\langle\bar{q}q\rangle({\rm 1 GeV})\left[\frac{\alpha_{s}({\rm 1 GeV})}{\alpha_{s}(\mu)}\right]^{\frac{12}{33-2n_f}}\, , \nonumber\\
 \langle\bar{s}s \rangle(\mu)&=&\langle\bar{s}s \rangle({\rm 1 GeV})\left[\frac{\alpha_{s}({\rm 1 GeV})}{\alpha_{s}(\mu)}\right]^{\frac{12}{33-2n_f}}\, , \nonumber\\
 \langle\bar{q}g_s \sigma Gq \rangle(\mu)&=&\langle\bar{q}g_s \sigma Gq \rangle({\rm 1 GeV})\left[\frac{\alpha_{s}({\rm 1 GeV})}{\alpha_{s}(\mu)}\right]^{\frac{2}{33-2n_f}}\, ,\nonumber\\
  \langle\bar{s}g_s \sigma Gs \rangle(\mu)&=&\langle\bar{s}g_s \sigma Gs \rangle({\rm 1 GeV})\left[\frac{\alpha_{s}({\rm 1 GeV})}{\alpha_{s}(\mu)}\right]^{\frac{2}{33-2n_f}}\, ,\nonumber\\
m_c(\mu)&=&m_c(m_c)\left[\frac{\alpha_{s}(\mu)}{\alpha_{s}(m_c)}\right]^{\frac{12}{33-2n_f}} \, ,\nonumber\\
m_s(\mu)&=&m_s({\rm 2GeV} )\left[\frac{\alpha_{s}(\mu)}{\alpha_{s}({\rm 2GeV})}\right]^{\frac{12}{33-2n_f}}\, ,\nonumber\\
\alpha_s(\mu)&=&\frac{1}{b_0t}\left[1-\frac{b_1}{b_0^2}\frac{\log t}{t} +\frac{b_1^2(\log^2{t}-\log{t}-1)+b_0b_2}{b_0^4t^2}\right]\, ,
\end{eqnarray}
  where $t=\log \frac{\mu^2}{\Lambda^2}$, $b_0=\frac{33-2n_f}{12\pi}$, $b_1=\frac{153-19n_f}{24\pi^2}$, $b_2=\frac{2857-\frac{5033}{9}n_f+\frac{325}{27}n_f^2}{128\pi^3}$,  $\Lambda=213\,\rm{MeV}$, $296\,\rm{MeV}$  and  $339\,\rm{MeV}$ for the flavors  $n_f=5$, $4$ and $3$, respectively  \cite{PDG}.
For the fully open flavor $cs\bar{u}\bar{d}$ tetraquark  states, we choose the flavor numbers $n_f=4$, and the typical energy scale $\mu=1\,\rm{GeV}$.

Let us choose the continuum threshold parameters  as  $\sqrt{s_0}=M_{X}+(0.5\sim0.7)\,\rm{GeV}=2.9+(0.5\sim0.7)\,\rm{GeV}$ tentatively according to the mass gap $m_{\psi^\prime}-m_{J/\psi}=0.59\,\rm{GeV}$ \cite{PDG}, and vary the parameters $\sqrt{s_0}$ to obtain   the best Borel parameters $T^2$ to satisfy  pole dominance at
 the hadron side and convergence of the operator product expansion at the QCD side.

 After trial and error, we obtain the ideal Borel parameters or Borel windows $T^2$ and continuum threshold parameters $s_0$, therefore  the pole contributions of the ground state scalar $cs\bar{u}\bar{d}$ tetraquark states and the convergent behaviors of the operator product expansion, see Table \ref{Borel}. In the Borel windows, the pole contributions are about $(38-67)\%$, while the central values exceed $52\%$, the pole dominance is well satisfied. The absolute values of the  contributions of the highest dimensional vacuum condensates $|D(11)|$ are about $(2-4)\%$ and $(0-1)\%$ for the  $AA$-type and $SS$-type tetraquark states, respectively. The operator product expansion is well convergent. At the beginning, we assume the ground states of the scalar tetraquark states $cs\bar{u}\bar{d}$  have the masses about $2.9\,\rm{GeV}$, just like the $X_0(2900)$,   and choose the continuum threshold parameters $\sqrt{s_0}=2.9+(0.5\sim0.7)\,\rm{GeV}$ tentatively to search for the optimal values via trial and error to satisfy the constraint  $\sqrt{s_0}=M_{X}+(0.5\sim0.7)\,\rm{GeV}$ besides the two basic criteria   of the QCD sum rules. From Table \ref{Borel}, we can see that for the $AA$-type scalar tetraquark state, the continuum threshold parameter $\sqrt{s_0}=2.9+(0.5\sim0.7)\,\rm{GeV}$ happens  to coincide  with the optimal value $3.5\pm 0.1 \,\rm{GeV}$, while for the  $SS$-type scalar tetraquark state, the continuum threshold parameter $\sqrt{s_0}=2.9+(0.5\sim0.7)\,\rm{GeV}$ is slightly smaller than  the optimal value $3.6\pm 0.1 \,\rm{GeV}$. In fact, we can choose other values of the continuum threshold parameters $\sqrt{s_0}$, for example, $\sqrt{s_0}=2.5+(0.5\sim0.7)\,\rm{GeV}$ as the initial point, and obtain the optimal values shown Table \ref{Borel}.

\begin{figure}
 \centering
 \includegraphics[totalheight=5cm,width=7cm]{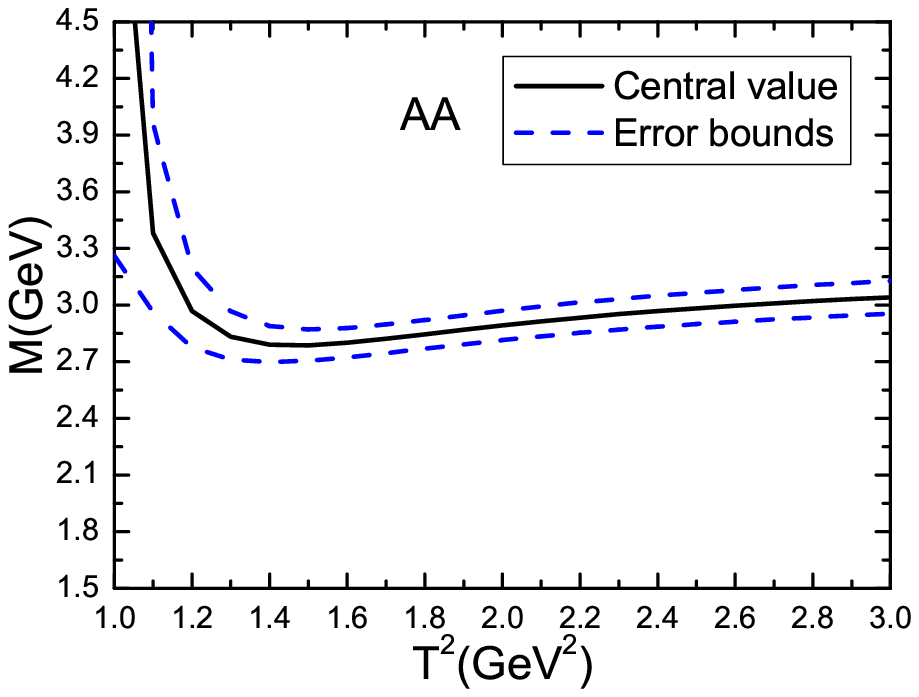}
  \includegraphics[totalheight=5cm,width=7cm]{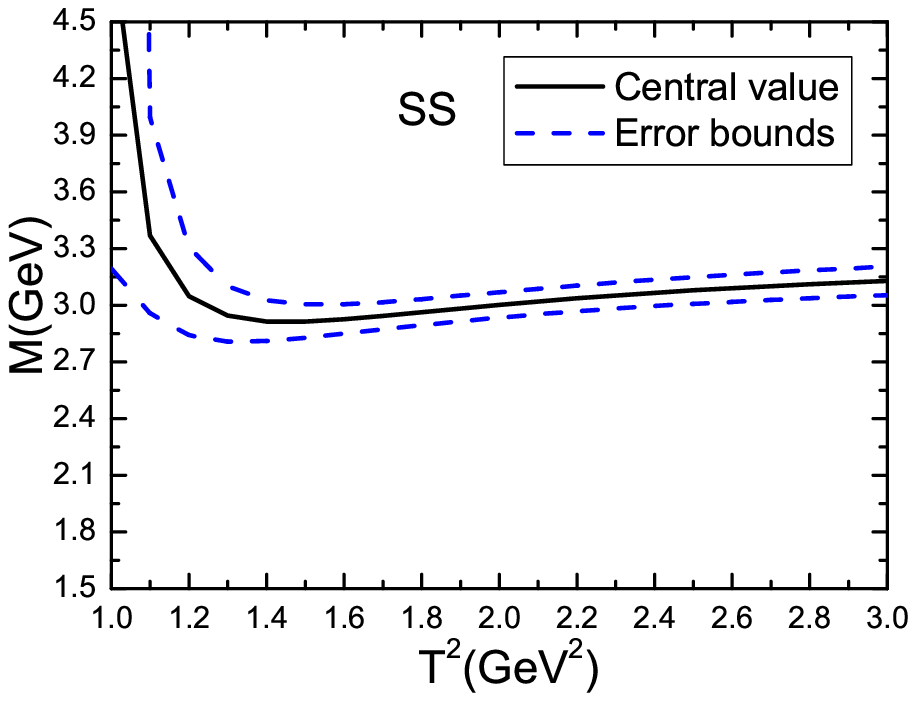}
 \caption{ The masses of the $AA$-type and $SS$-type tetraquark states with variations  of the Borel parameters $T^2$.   }\label{mass-fig}
\end{figure}

Now we take  into account all uncertainties of the input parameters, and obtain the values of the masses and pole residues of
 the fully  open flavor $cs\bar{u}\bar{d}$ tetraquark  states, which are  shown explicitly in Table \ref{Borel} and Fig.\ref{mass-fig}.
 In Fig.\ref{mass-fig}, we plot the masses  of the  $AA$-type and $SS$-type scalar $cs\bar{u}\bar{d}$ tetraquark states with variations  of the Borel parameters $T^2$ in  much larger  ranges than the Borel windows.
 From the figure, we can see that  there appear  platforms in the Borel windows,   it is reliable to extract the tetraquark masses.

The predicted mass $M_{AA}=2.91\pm0.12\,\rm{GeV}$ is consistent with the experimental value $2866\pm7~{\rm MeV}$ from the LHCb collaboration \cite{LHCbTalk:2020-1,LHCbTalk:2020-2}, and supports assigning the $X_0(2900)$ to be the $AA$-type  $cs\bar{u}\bar{d}$ tetraquark state with the spin-parity $J^P=0^+$. While the predicted mass $M_{SS}=3.05\pm0.10\,\rm{GeV}$ lies above the experimental value $2866\pm7~{\rm MeV}$ from the LHCb collaboration \cite{LHCbTalk:2020-1,LHCbTalk:2020-2}.

The two-body strong decays $X_0(2900)\to D\bar{K}$ can take place with the fall-apart mechanism and are kinematically allowed, therefore it is Okubo-Zweig-Iizuka super-allowed. The current $J_{AA}(x)$ also couples potentially to the two-meson scattering states $D\bar{K}$, which leads  to a finite width to the $X_0(2900)$. The experimental value $\Gamma_{X_0}=57\pm13~{\rm MeV}$ is small enough, the finite width effect can be neglected safely.
Analogous decay widths are obtained for the charmed partners $[su][\bar{c}\bar{d}]$ of the $X(5568)$ \cite{KAzizi}.
In Ref.\cite{Wang-IJMPA-Zc3900}, we study the $Z_c(3900)$ with the QCD sum rules in details by including the two-particle scattering state contributions   and nonlocal effects between the diquark and antidiquark constituents. The two-particle scattering state contributions, such as the $J/\psi\pi$, $\eta_c \rho$, $D\bar{D}^*$+$h.c.$, etc,  cannot saturate  the QCD sum rules at the hadron side,  the contribution of the $Z_c(3900)$ plays an un-substitutable role, we can saturate the QCD sum rules with or without the two-particle scattering state contributions. The conclusion is applicable in the present case.

 The contributions of the intermediate   two-meson  scattering states  $D\bar{K}$, $D^* \bar{K}^*$,  etc besides the scalar tetraquark candidate $X_0(2900)$  can be written as,
\begin{eqnarray}\label{Self-Energy}
\Pi_{AA}(p^2) &=&-\frac{\widehat{\lambda}_{X}^{2}}{ p^2-\widehat{M}_{X}^2+\Sigma_{D\bar{K}}(p^2)+\Sigma_{D^* \bar{K}^*}(p^2)+\cdots}+\cdots \, .
\end{eqnarray}
 We choose the bare mass and pole residue $\widehat{M}_{X}$ and $\widehat{\lambda}_{X}$  to absorb the divergences in the self-energies $\Sigma_{D\bar{K}}(p^2)$, $\Sigma_{D^* \bar{K}^*}(p^2)$, etc.
The renormalized self-energies  contribute  a finite imaginary part to modify the dispersion relation,
\begin{eqnarray}
\Pi_{AA}(p^2) &=&-\frac{\lambda_{X}^{2}}{ p^2-M_{X}^2+i\sqrt{p^2}\,\Gamma_{X}(p^2)}+\cdots \, ,
 \end{eqnarray}
with  the (central value of) physical  width $\Gamma_{X}(M_X^2)=57 \, \rm{MeV}$ from  the LHCb collaboration \cite{LHCbTalk:2020-1,LHCbTalk:2020-2}.

We can take into account the finite width  with  the simple replacement of the hadronic spectral density,
\begin{eqnarray}
\lambda^2_{X}\delta \left(s-M^2_{X} \right) &\to& \lambda^2_{X}\frac{1}{\pi}\frac{M_{X}\Gamma_{X}(s)}{(s-M_{X}^2)^2+M_{X}^2\Gamma_{X}^2(s)}\, ,
\end{eqnarray}
where
\begin{eqnarray}
\Gamma_{X}(s)&=&\Gamma_{X} \frac{M_{X}^2}{s}\sqrt{\frac{s-(M_{D}+M_{K})^2}{M^2_{X}-(M_{D}+M_{K})^2}} \, .
\end{eqnarray}
Then the hadron  sides of  the QCD sum rules in Eqs.\eqref{QCDSR}-\eqref{QCDSR-mass} undergo the replacements,
\begin{eqnarray}
\lambda^2_{X}\exp \left(-\frac{M^2_{X}}{T^2} \right) &\to& \lambda^2_{X}\int_{(m_{D}+m_{K})^2}^{s_0}ds\frac{1}{\pi}\frac{M_{X}\Gamma_{X}(s)}{(s-M_{X}^2)^2+M_{X}^2\Gamma_{X}^2(s)}\exp \left(-\frac{s}{T^2} \right)\, , \nonumber\\
&=&(0.97\sim0.98)\,\lambda^2_{X}\exp \left(-\frac{M^2_{X}}{T^2} \right)\, , \\
\lambda^2_{X}M^2_{X}\exp \left(-\frac{M^2_{X}}{T^2} \right) &\to& \lambda^2_{X}\int_{(m_{D}+m_{K})^2}^{s_0}ds\,s\,\frac{1}{\pi}\frac{M_{X}\Gamma_{X}(s)}{(s-M_{X}^2)^2+M_{X}^2\Gamma_{X}^2(s)}\exp \left(-\frac{s}{T^2} \right)\, , \nonumber\\
&=&(0.96\sim0.96)\,\lambda^2_{X}M^2_{X}\exp \left(-\frac{M^2_{X}}{T^2} \right)\, ,
\end{eqnarray}
with the central value of the continuum threshold parameter  $\sqrt{s_0}=3.50\,\rm{GeV}$.
We can absorb the numerical factors  $0.97\sim0.98$ and $0.96\sim0.96$ into the pole residue with the simple replacement $\lambda_{X}\to (0.98\sim0.99)\lambda_X$ safely. It is indeed that the finite width effects  cannot  affect  the mass $M_{X}$ and pole residue $\lambda_{X}$ remarkably.
However, we should bear in mind that there are non-pole contributions from the two-meson scattering states besides modifying the dispersion relation, which are expected to play a minor important role in the vicinity of the pole.

\begin{table}
\begin{center}
\begin{tabular}{|c|c|c|c|c|c|c|c|}\hline\hline
                              &$T^2(\rm{GeV}^2)$   &$\sqrt{s_0}(\rm{GeV})$  &pole         &$|D(11)|$  &$M(\rm{GeV})$  &$\lambda(\rm{GeV}^5)$ \\ \hline

$[cs]_A[\bar{u}\bar{d}]_{A}$  &$1.9-2.3$           &$3.5\pm0.1$             &$(38-67)\%$  &$(2-4)\%$  &$2.91\pm0.12$  &$(1.60\pm0.33)\times10^{-2}$   \\ \hline

$[cs]_S[\bar{u}\bar{d]}_{S}$  &$2.1-2.5$           &$3.6\pm0.1$             &$(39-66)\%$  &$(0-1)\%$  &$3.05\pm0.10$  &$(1.20\pm0.21)\times10^{-2}$   \\ \hline\hline
\end{tabular}
\end{center}
\caption{ The Borel windows, continuum threshold parameters, pole contributions, contributions of the vacuum condensates of dimension $11$,  masses and pole residues for the scalar  tetraquark states. } \label{Borel}
\end{table}

\section{Conclusion}
In this article, we construct the axialvector-diquark-axialvector-antidiquark type and scalar-diquark-scalar-antidiquark type currents to study the fully open flavor $cs\bar{u}\bar{d}$ tetraquark states with the spin-parity $J^P={0}^+$ via the QCD sum rules by carrying  out the operator product expansion up to the vacuum condensates of dimension 11 in a consistent way. We obtain the predictions $M_{AA}=2.91\pm0.12\,\rm{GeV}$ and $M_{SS}=3.05\pm0.10\,\rm{GeV}$, the predicted mass for the axialvector-diquark-axialvector-antidiquark type scalar tetraquark state is consistent with the experimental value $2866\pm7~{\rm MeV}$ from the LHCb collaboration, and supports assigning the $X_0(2900)$ to be the axialvector-diquark-axialvector-antidiquark type  $cs\bar{u}\bar{d}$ tetraquark state with the spin-parity $J^P=0^+$.

\section*{Acknowledgements}
This  work is supported by National Natural Science Foundation, Grant Number  11775079.

\end{document}